\documentclass{ws-mpla-mod}
\usepackage{graphicx}
\usepackage[usenames,dvipsnames]{color}
\usepackage{mathrsfs} 
\usepackage{soul}

\newcommand{\gev}{~\mathrm{GeV}}

\newcommand{\oblique}{Altarelli:1990zd,Peskin:1990zt,Peskin:1991sw,Maksymyk:1993zm}

\newcommand{\lb}{\left (}
\newcommand{\rb}{\right )}
\newcommand{\al}{\alpha}
\newcommand{\be}{\beta}
\newcommand{\eqn}{equation}
\newcommand{\GeV}{{\ensuremath\rm GeV}}
\newcommand{\MeV}{{\ensuremath\rm MeV}}
\newcommand{\TeV}{{\ensuremath\rm TeV}}
\newcommand{\pb}{{\ensuremath\rm pb}}

\newcommand{\lam}{\lambda}

\begin{document}
\bibliographystyle{unsrt}
\markboth{{Agnieszka Ilnicka}, Tania Robens, {Tim Stefaniak}}
{Constraining extended scalar sectors at the LHC and beyond}

\catchline{}{}{}{}{}

\title{Constraining Extended Scalar Sectors at the LHC and beyond}

\author{\footnotesize Agnieszka Ilnicka}

\address{Institute of Physics, University of Zurich\\
Winterthurstrasse 190, CH-8057 Zurich, Switzerland\\
  ailnicka@physik.uzh.ch}

\author{\footnotesize Tania Robens}

\address{MTA-DE Particle Physics Research Group, University of Debrecen\\ 4010 Debrecen, Hungary\\
  tania.robens@science.unideb.hu}

\author{\footnotesize Tim Stefaniak}

\address{Deutsches Elektronen-Synchrotron DESY\\
Notkestra{\ss}e 85, 22607 Hamburg, Germany\\
tim.stefaniak@desy.de}

\maketitle


\begin{abstract}
We give a brief overview of beyond the Standard Model (BSM) theories with an extended scalar sector and their phenomenological status in the light of recent experimental results. We discuss the relevant theoretical and experimental constrains, and show their impact on the allowed parameter space of two specific models: the real scalar singlet extension of the Standard Model (SM) and the Inert Doublet Model. We emphasize the importance of the LHC measurements, both the direct searches for additional scalar bosons, as well as the precise measurements of properties of the Higgs boson of mass 125 GeV. We show the complementarity of these measurements to electroweak and dark matter  observables.

\end{abstract}

\ccode{PACS Nos.: 12.60.Fr, 14.80.Ec, 14.80.Fd}

\ccode{Preprint number: DESY 18-031}

\section{Introduction}

With the startup of data taking at the LHC in 2010 particle physics has entered an exciting era, with a milestone reached in 2012 with the discovery of a Higgs boson with a mass of $125\gev$.\cite{Aad:2012tfa,Chatrchyan:2012xdj} The experimental results from the ATLAS and CMS collaborations from Run I and the currently ongoing Run II are in good agreement with the predictions of the Standard Model, in particular, the discovered Higgs particle appears to be consistent with the expectations for a SM Higgs boson.\cite{Khachatryan:2016vau} However, both the experimental and theoretical uncertainties still leave room for new physics.

Extensions of the scalar sector can provide intriguing scenarios for new physics. Additional terms in the scalar potential may elicit a strong first-order electroweak phase transition \cite{Bochkarev:1990fx,Turok:1990zg,Land:1992sm,Cline:1995dg,Cline:1996mga,Lahanas:1998wf}, 
and/or provide new sources of CP violation {(see e.g. Ref.~\refcite{Branco:2011iw})}
--- both are needed if the matter-antimatter asymmetry of the Universe is generated during the electroweak phase transition.\cite{Morrissey:2012db,Konstandin:2013caa} Moreover, models with extended scalar sectors may feature a suitable candidate for the dark matter (DM) observed in our Universe and/or scalar particle(s) that mediate the interactions of the visible to a dark sector (i.e.~so-called \emph{Higgs portal} models).\cite{Patt:2006fw} 
 From the UV perspective, models with extended Higgs sectors are often obtained as an effective low-energy description of a more complete BSM theory (e.g.~Supersymmetry), which in turn may address other shortcomings of the SM, e.g.~the hierarchy problem or the unification of gauge couplings.\cite{Gunion:1989we}

In this review, we discuss important phenomenological constraints that generally need to be taken into account in the investigation of models with an extended scalar sector.\footnote{Naturally, in this brief review we cannot give a complete overview of this broad topic. We therefore confine ourselves to two simple, illustrative BSM scenarios. Please see Ref. \refcite{Ivanov:2017dad} for a more extensive recent review.}
 We demonstrate these constraints on two simple models, in which the SM Higgs sector is extended by \emph{(i)} a real scalar singlet field, and \emph{(ii)} an inert scalar $\mathrm{SU}(2)_L$ doublet. The first model can be considered as a minimal extension of the SM with only one additional field, whereas the second model is very attractive as the lightest inert scalar boson in the model is a suitable DM candidate.  A full discussion of all constraints in these models has been presented in Refs.~\refcite{Pruna:2013bma,Lopez-Val:2014jva,Robens:2015gla,Ilnicka:2015jba,Robens:2016xkb,Ilnicka:2017gab}. Here, we only emphasize the most important constraints and refer the reader to the above references for a complete study.
 
\subsection{$(i)$ Higgs singlet extension}
{In} the \emph{Higgs singlet extension}, in its simplest version, the SM scalar sector {is augmented} {by a} {real scalar field} that is a singlet under the SM gauge group. {The scalar potential can be simplified} by requiring the model to be renormalizable and imposing additional symmetries. We here discuss the case where the model is symmetric under an additional $Z_2$ symmetry.\cite{Patt:2006fw,Schabinger:2005ei}

Under these assumptions the scalar potential is given by 
\begin{eqnarray}\label{potential}
V(\Phi,S ) 
= -m^2 \Phi^{\dagger} \Phi -\mu ^2 S ^2 + \lambda_1
(\Phi^{\dagger} \Phi)^2 + \lambda_2  S^4 + \lambda_3 \Phi^{\dagger}
\Phi S ^2,
\end{eqnarray}
with the two scalar fields given by
\begin{equation}\label{unit_higgs}
\Phi \equiv \frac{1}{\sqrt{2}}
\left(
\begin{gathered}
0 \\
\tilde{h}+v
\end{gathered} \right), 
\hspace{2cm}
S \equiv \frac{h'+x}{\sqrt{2}},
\end{equation} 
in the unitary gauge, and $\tilde{h}$ and $h'$ being the dynamical degrees of freedom and $v$ and $x$ the vacuum expectation values (vevs). The kinetic term of the Lagrangian reads
\begin{eqnarray}
\mathscr{L}_{s,\text{kin}} = \left( D^{\mu} \Phi \right)^{\dagger} D_{\mu} \Phi +  \partial^{\mu} S \partial_{\mu} S.
\end{eqnarray}
{The gauge eigenstates are rotated into the mass eigenstates by a mixing angle $\al$,
\begin{eqnarray}
\left( \begin{array}{c} h \\ H \end{array} \right ) = \begin{pmatrix} \cos\al & -\sin\al \\ \sin\al & \cos\al \end{pmatrix} \left( \begin{array}{c} \tilde{h} \\ h' \end{array} \right).
\end{eqnarray}
}
Another important parameter is the ratio of the two vevs, $\tan\be \equiv v/x$. After minimization of the scalar potential, the model contains five free parameters, which can be chosen to be
\begin{\eqn*}
m_h,\,m_H,\,\sin\al,\,\tan\be,\,v,
\end{\eqn*}
where $m_h\,(m_H)$ denotes the mass of the lighter (heavier) mass eigenstate. Electroweak precision observables set $v\sim246\gev$. One of the two Higgs particles, $h$ or $H$, must be identified with the Higgs particle measured by the LHC experiments with a mass of $\sim125\gev$. This leaves three free parameters, which are further constrained by theoretical and experimental constraints, as discussed below.

\subsection{$(ii)$ Inert Doublet Model}
Another intriguing extension of the SM scalar sector {is the \emph{Inert Doublet Model} (IDM)\cite{Deshpande:1977rw,Cao:2007rm,Barbieri:2006dq} which features a dark matter candidate. In this two Higgs doublet model, the additional $\mathrm{SU}(2)_L$ doublet, {$\phi_D$}, is made \emph{inert} for the SM matter sector by imposing a discrete {$Z_2$} symmetry, which we call $ D$-symmetry, {with} the following transformation properties:
\begin{eqnarray}
\phi_S\to \phi_S, \quad \phi_D \to - \phi_D, \quad
\text{SM~field} \to \text{SM~field}.
\end{eqnarray}
This discrete symmetry is respected by the Lagrangian and the vacuum{, i.e.~$\phi_D$ does not acquire a vev}. {The other doublet, $\phi_S$, on the other hand,} plays the same role {as the Higgs doublet in the SM, and yields a SM-like Higgs state, $h$}. {The} so-called \emph{inert} or \emph{dark} doublet {$\phi_D$} contains {four scalar degrees of freedom that manifest as a charged scalar boson pair, $H^\pm$, and two neutral scalar bosons, $H$ and $A$,} with the {lighter neutral inert scalar} being a natural DM candidate. The potential of this model is given by
\begin{equation}\begin{array}{c}
V(\phi_S,\phi_D)=-\frac{1}{2}\left[m_{11}^2(\phi_S^\dagger\phi_S)\!+\! m_{22}^2(\phi_D^\dagger\phi_D)\right]+
\frac{\lambda_1}{2}(\phi_S^\dagger\phi_S)^2\! 
+\!\frac{\lambda_2}{2}(\phi_D^\dagger\phi_D)^2\\[2mm]+\!\lambda_3(\phi_S^\dagger\phi_S)(\phi_D^\dagger\phi_D)\!
\!+\!\lambda_4(\phi_S^\dagger\phi_D)(\phi_D^\dagger\phi_S) +\frac{\lambda_5}{2}\left[(\phi_S^\dagger\phi_D)^2\!
+\!(\phi_D^\dagger\phi_S)^2\right].
\end{array}\label{pot}\end{equation}
After the minimization of the potential, the model has in total seven a priori free parameters
\begin{\eqn}\label{eq:idm_pars}
\underbrace{M_h,\;v}_{\phi_S},\; \underbrace{M_H,\; M_A,\; M_{H^{\pm}}}_{\phi_D},\; \underbrace{\lam_2,\; \lam_{345}}_{\text{IDM}},
\end{\eqn}
{that are associated to the SM-like scalar doublet, $\phi_S$, the dark scalar doublet, $\phi_D$, or the scalar potential, Eq.~\eqref{pot}, respectively.}
 Due to the different transformation properties under the $D$-symmetry, there is no mixing of the neutral scalar fields between the two doublets. The coupling $\lam_2$ {governs the} interactions within the dark sector, while $\lam_{345}\,\equiv\,\lam_3+\lam_4+\lam_5$ is responsible for the interactions between the SM-like Higgs boson $h$ and the {DM particle}. The {inert} doublet, $\phi_D$, transforms equivalently to $\phi_S$ under the SM gauge group; however, due to the exact $Z_2$ symmetry, it does not contribute to electroweak symmetry breaking. In turn, this implies that both $v$ and $M_h$ are fixed by electroweak precision observables and the mass measurement of the Higgs boson observed at the LHC. This leaves in total 5 undetermined parameters, {which are subject to various theoretical and experimental constraints (see below)}. Choosing one of the new scalars as the DM candidate naturally implies a mass hierarchy {in the inert sector}. Here we chose $H$ as DM candidate, hence $M_H\,\leq\,M_A,\,M_{H^\pm}$.\footnote{Note that choosing $A$ as DM candidate, i.e.~$M_A \le M_H, M_{H^\pm}$, is phenomenologically identical to our choice, {with the replacement} $\lam_{345}\,\longleftrightarrow\,\bar{\lam}_{345}\,\equiv\lam_3+\lam_4-\lam_5$ everywhere.}}

\section{Theoretical constraints} 
In general, all extensions of the SM are subject to a large number of theoretical and experimental constraints. {Quite generally,} the following conditions {have to be fulfilled in order to have a reliable perturbative description of the model}:
\begin{romanlist}[(iii)]
\item{}the potential must be bounded from below and display at least a local (if not global) minimum;
\item{}the matrix of all $2\,\rightarrow\,2$ scattering processes must be unitary; this condition is imposed by making use of perturbative unitarity\cite{Lee:1977eg}, which renders an upper limit on the coefficients of the partial wave expansion for the scattering matrix;
\item{}perturbativity of the couplings; for a coupling $\lam$, this implies $|\lam|\,\leq\,4\pi$.
\end{romanlist}
Conditions (i) and (iii) above directly lead to constraints on (relations of) couplings; the second condition depends on more involved relations, but can easily be implemented using numerical tools.

For the IDM, two {minima} can coexist, leading to an additional requirement in order to guarantee the {{inert}} vacuum to be global.\cite{Ginzburg:2010wa,Gustafsson:2010zz,Swiezewska:2012ej}\footnote{See Ref. \refcite{Khan:2015ipa} for a detailed study regarding metastable vacua within the IDM.}

\section{Experimental constraints from collider experiments}

A BSM theory with an extended scalar sector can only be phenomenologically viable if \emph{(i)} it features a Higgs boson that is consistent with the observed properties of the Higgs boson signal at the LHC, and \emph{(ii)} the remaining scalar states of the model are not in conflict with the null-results in collider searches for additional scalar states. For the latter, besides the current collider experiment (LHC) it can also be important to take limits from past collider experiments (LEP, Tevatron) into account. It is then an interesting question how these collider constraints compare to other, complementary probes for BSM effects, and what the future prospects are for a discovery/exclusion of the model.

\subsection{Searches for additional Higgs bosons}

In the real singlet extension, the couplings of the additional Higgs state to SM particles are reduced universally by the mixing angle; assuming that the lighter Higgs boson $h$ is the discovered SM-like Higgs boson with $m_h \simeq 125\gev$, the couplings of $H$ to SM bosons and fermions are rescaled by $\sin\alpha$. In addition, if $m_H > 2 m_h$, the additional decay mode $H\to hh$ is kinematically allowed and can be sizable. Hence, two types of LHC searches are relevant, namely, for direct Higgs production $pp\to H$ with successive decay to SM particles (mostly importantly, $WW$ and $ZZ$)\cite{CMS-PAS-HIG-12-045,CMS-PAS-HIG-13-003,Khachatryan:2015cwa,CMS:17012,Aaboud:2017rel} or to a pair of SM-like Higgs bosons\cite{Aad:2014yja,Khachatryan:2016sey}. Constraints from these (and other) Higgs searches can be conveniently tested for (nearly) arbitrary BSM theories with the public tool \texttt{HiggsBounds}\cite{Bechtle:2008jh,Bechtle:2011sb,Bechtle:2013gu,Bechtle:2013wla,Bechtle:2015pma}, which we  also employ here.

\begin{figure}[htb!]
\centerline{\includegraphics[width=0.8\textwidth]{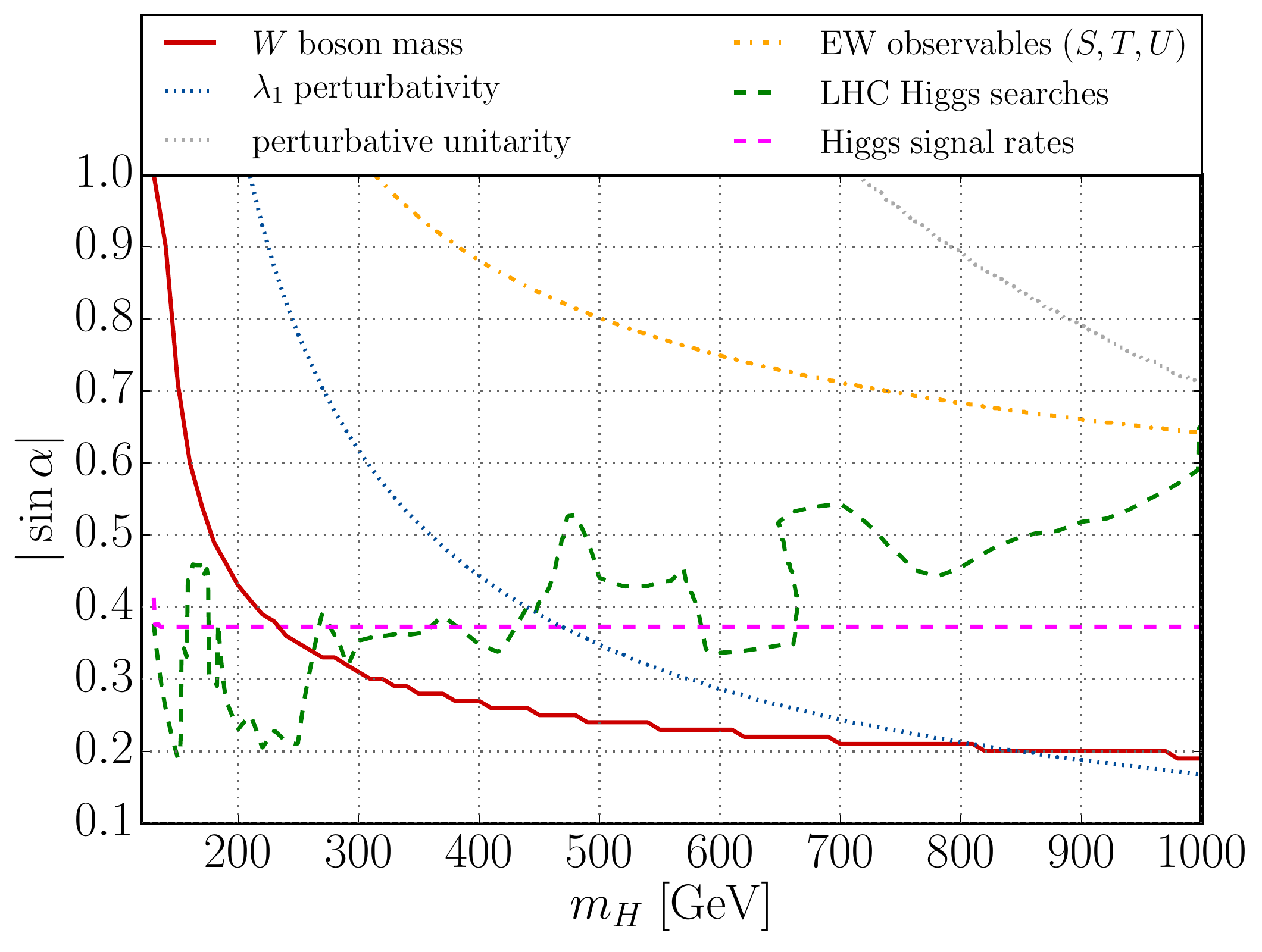}}
\vspace*{8pt}
\caption{Constraints on the mixing angle, $|\sin\al|$, as a function of the additional Higgs boson mass, $m_H$, in the real singlet extension of the SM (for fixed $\tan\beta = 0.1$): Theoretical constraints from perturbativity of the couplings (\emph{dotted blue}), perturbative unitarity (\emph{dotted gray}), and experimental constraints from the $W$-boson mass (\emph{solid red}), electroweak precision observables (\emph{dot-dashed orange}), LHC Higgs searches for additional Higgs bosons (\emph{dashed green}) and LHC measurements of the Higgs signal rates (\emph{dashed magenta}). 
\protect\label{fig:singlet_constraints}}
\end{figure}

In Fig.~\ref{fig:singlet_constraints}, we show a comparison of theoretical and experimental constraints on the parameters $m_H$ and $\sin\alpha$ in the real singlet extension (for fixed $\tan\beta=0.1$). Current LHC searches for additional Higgs bosons (\emph{green dotted line}) provide the strongest constraint in the lower mass range (i.e., roughly, $m_H\in [130,~300]$). The displayed constraints are obtained from Run~II ATLAS and CMS searches with $WW$ and $ZZ$ final states using $\sim 36~\mathrm{fb}^{-1}$ of collected data.\cite{CMS:aya,CMS:bxa,Khachatryan:2015cwa,CMS:17012,Aaboud:2017rel} At higher masses, indirect constraints obtained from the $W$-boson mass (see Sect.~\ref{Sect:EWPOs}) are more constraining, {however, could also be alleviated more easily by new physics \emph{beyond} the real singlet extension}. Yet, the LHC direct searches remain to provide useful and robust constraints on the mixing angle even for larger $m_H$.

For the IDM, on the other hand, direct production of the additional (inert) scalar states is suppressed due to the $Z_2$ symmetry, which only allows them to be produced pair-wise. The conventional LHC searches for additional Higgs states thus do not apply straight-forwardly to the IDM.
However, results from LEP and LHC searches for supersymmetric particles can be recasted to the IDM, yielding strong constraints on certain regions of parameter space.
A reinterpretation of a LEP search for neutralino pair production\cite{EspiritoSanto:2003by} excluded a region of the IDM parameter space where
\begin{\eqn}\label{eq:leprec}
M_A\leq100\gev,\quad M_H\leq80\gev\quad\text{and}\quad \Delta M {(A,H)}\geq8\gev,
\end{\eqn}
simultaneously.\cite{Lundstrom:2008ai} This limit is taken into account in the results presented here.\cite{Ilnicka:2015jba}

At the LHC the searches for final states containing multi-leptons and missing transverse energy show the best sensitivity to the IDM.\cite{Belanger:2015kga} However, Ref.~\refcite{Ilnicka:2015jba} pointed out that the regions excluded by these searches are also subject to other constraints, e.g.~from DM detection and the observed DM relic abundance.

{The most promising LHC signatures in the IDM} contain final states with a single or two electroweak gauge bosons, {associated with} a pair of DM particles which lead to  missing transverse energy. Collider studies at the LHC for the IDM have e.g. been presented in Refs. \refcite{Cao:2007rm,Dolle:2009ft,Miao:2010rg,Gustafsson:2012aj,Blinov:2015qva,Poulose:2016lvz,Kanemura:2016sos,Akeroyd:2016ymd,Datta:2016nfz,Hashemi:2016wup}. Many supersymmetric searches {target} the same final states and are therefore ideal candidates for reinterpretation. However, the decay topologies typically differ in these two models. The {event selection} of supersymmetric searches is therefore not optimized for {similar processes in the IDM. Hence, future experimental searches designed specifically to the IDM processes would significantly enhance the LHC discovery potential for the IDM.}

\subsection{Limits from Higgs signal rates}
\label{sect:higgssignal}

BSM effects can modify the signal rates of the $125\gev$ Higgs boson with respect to the SM prediction in various ways, e.g.~
\begin{itemize}
\item by directly modifying its tree-level couplings to SM particles;
\item by modifying its couplings to SM particles at the loop-level. This is particularly relevant in cases where the leading SM and BSM contributions enter at the same order of perturbation theory, which is often the case for the Higgs couplings to photons and gluons;
\item by introducing genuine Higgs decay channels that do not exist for the SM Higgs boson. Generically, new decay mode(s) lead to an overall suppression of the Higgs decays to SM final states.
\end{itemize}
Hence, quite generally, the LHC measurements of the Higgs signal rates provide important constraints on the BSM parameter space. These can be easily evaluated for (nearly) arbitrary BSM theories with the public code \texttt{HiggsSignals}~\cite{Bechtle:2013xfa} (see also Ref.~\refcite{Bechtle:2014ewa}) by means of a $\chi^2$ test.
 
The two models considered here nicely demonstrate the three possible Higgs rate modifications. In the real singlet extension, the  tree-level Higgs couplings to SM particles are uniformly rescaled by the mixing angle (either by $\cos\alpha$ or $\sin\alpha$, depending on whether $h$ or $H$ is the Higgs candidate at $125\gev$). In addition, if $2 m_h \le m_H \simeq 125\gev$, the decay channel $H\to hh$ is kinematically allowed. In contrast, in the IDM, the tree-level Higgs couplings are unmodified, however, the charged scalar boson $H^\pm$ can contribute to the Higgs-photon-photon coupling at one-loop level\cite{Arhrib:2012ia,Swiezewska:2012eh} (i.e., at the same order as SM contributions). Furthermore, additional decay modes of the Higgs candidate at $125\gev$ to pairs of the inert scalars can be possible.

 We employed \texttt{HiggsSignals} to check the Higgs signal rate constraints in the real scalar singlet extension, using the results from the ATLAS and CMS Run~I combination\cite{Khachatryan:2016vau} and recent Run~II results\cite{ATLAS:2016gld,ATLAS:2016oum,ATLAS:2016pkl,ATLAS:2016awy,ATLAS:2016nke,ATLAS:2016ldo,Sirunyan:2017exp,CMS:2016ixj}. Here, we only focus on the mass region $m_H \gtrsim 130\gev$, with the lighter Higgs boson being the Higgs candidate at $125\gev$, {where we obtain an upper limit of $|\sin\al| \lesssim 0.37$ at $95\%$ confidence level (C.L.), see Fig.~\ref{fig:singlet_constraints}}. For the lower mass range, and in particular the inverted case with $H$ at $125\gev$, it is important to take into account a potential signal overlap of both Higgs bosons, as well as the additional decay mode $H\to hh$.\footnote{These cases are discussed in detail in Ref.~\refcite{Robens:2015gla}. Note also that for the parameter constellation $\tan\be=-\cot\al$ the decay rate for $H\to hh$ vanishes independent of the masses.}

\begin{figure}[tb!]
\begin{center}
\includegraphics[width=0.49\textwidth]{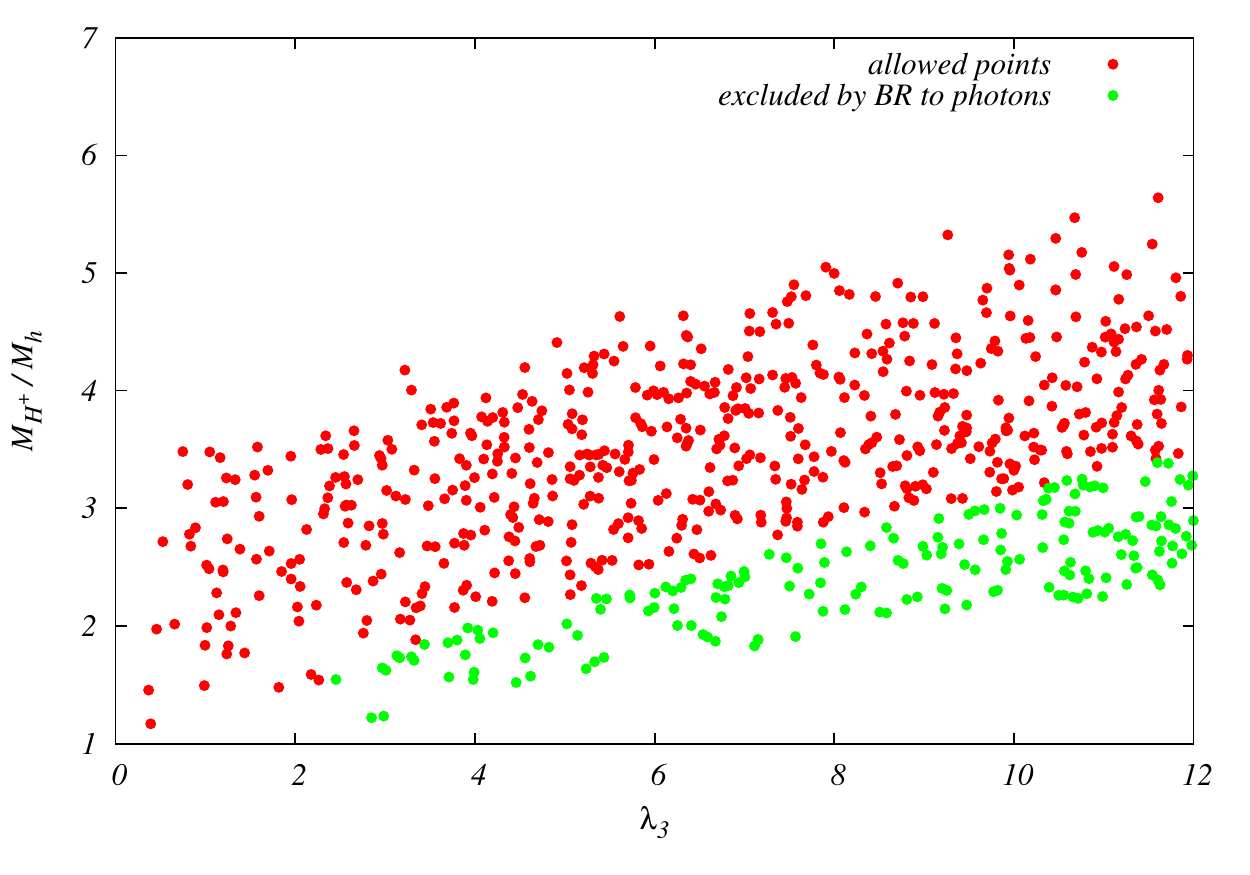}
\includegraphics[width=0.49\textwidth]{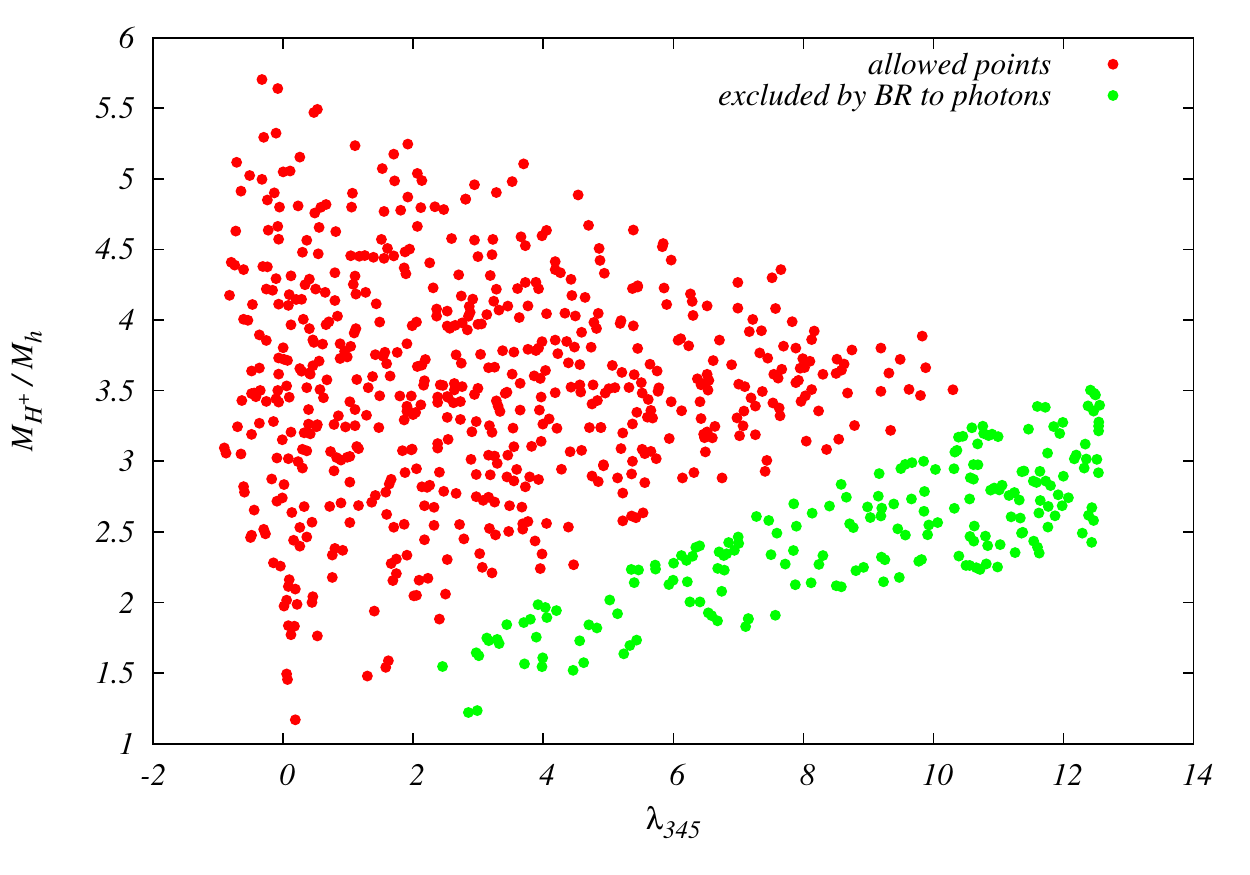}
\caption{\label{fig:digamma} {Constraints on the Inert Doublet Model from the LHC Higgs signal rate for $h\,\rightarrow\,\gamma\gamma$}: Points allowed (red) and excluded (green) in the $(\lam_3,M_{H^\pm}/M_h)$ {\sl (left)} and $(\lam_{345},M_{H^\pm}/M_h)$ {\sl (right)} plane. Here we allow for {under-abundant} DM{, such that large $\lam_{345}$ values are possible.} This plot only contains points with DM masses $m_H\,\geq\,130\,\GeV$. Figure taken from Ref.~\protect\refcite{Ilnicka:2017gab}.}
\end{center}
\end{figure}

For the Inert Doublet Model the numerical results presented here\cite{Ilnicka:2015jba,Ilnicka:2017gab} employ the public code \texttt{2HDMC}\cite{Eriksson:2009ws} to obtain the Higgs sector predictions. In particular, this includes the branching ratio predictions for the invisible Higgs decay, $\mathrm{BR}(h\to HH)$, and the loop-induced Higgs to diphoton decay, $\mathrm{BR}(H\to \gamma\gamma)$, which receives contributions from the inert charged scalar boson. The parameters relevant for this loop-induced process are $\lam_3$ and the ratio $M_{H^+}/M_h$, which are shown in Fig.~\ref{fig:digamma} (\emph{left}). As can be seen, the Higgs rate measurements yield a clear exclusion in the lower right corner, i.e.~for large $\lam_3$ and small ratios $M_{H^+}/M_h$. {Similarly, due to the linear connection between $\lambda_3$ and $\lambda_{345}$, we obtain a clear discrimination between allowed and excluded regions in the input parameter plane $\lb \lam_{345},\; M_{H^+}/M_h\rb$,  see Fig.~\ref{fig:digamma} (\emph{right}).}

\begin{figure}[htb!]
\begin{center}
\includegraphics[width=0.8\textwidth]{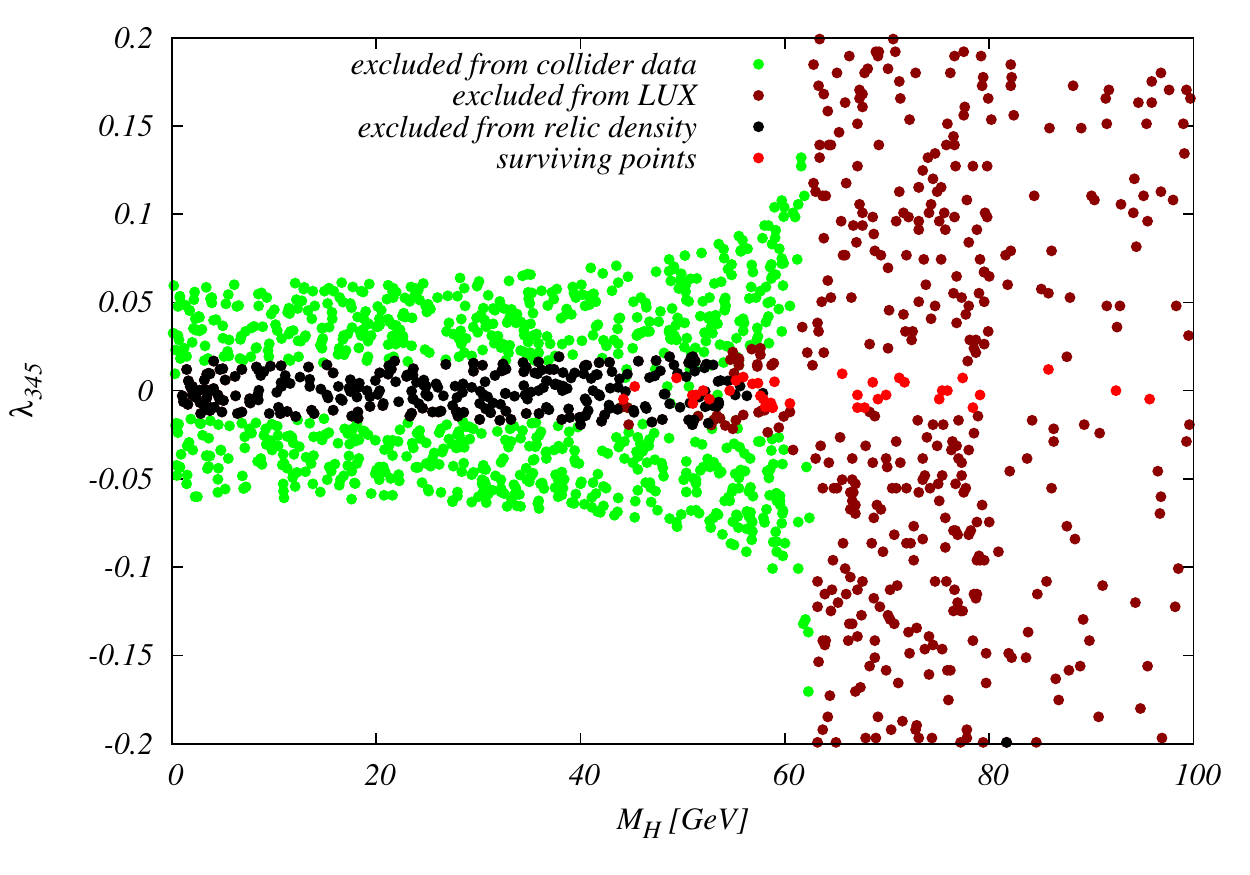}
\caption{\label{fig:lowmhidm_luxii} {Comparison of constraints in the Inert Doublet Model in the $(M_H,\lam_{345})$ plane for low DM masses: We show parameter points excluded by collider data, i.e.~the limit on the invisible decay rate of the SM-like Higgs boson inferred from the ATLAS and CMS combined Run~1 results in \emph{green}. Direct DM detection searches and DM relic abundance (see Sect.~\ref{sect:DM}) exclude the points shown in \emph{brown} and \emph{black}, respectively. The points surviving all constraints are shown in \emph{red}.} Figure taken from Ref.~\protect\refcite{Ilnicka:2017gab}.}
\end{center}
\end{figure}

The parameter region with $M_H\,\leq\,M_h/2$ is very sensitive to branching ratio limits for the invisible decay of the SM-like Higgs boson. This constraint alone already yields the limit $|\lam_{345}| \lesssim\,0.02$ in this mass range, see Fig.~\ref{fig:lowmhidm_luxii}. In this result we used rate measurements from the ATLAS and CMS Run~I combination.\cite{Khachatryan:2016vau}

\section{Other constraints}
Besides the collider constraints from Higgs searches and measurements, as discussed above, both models are subject to additional constraints. The most important ones, namely the constraints from electroweak precision observables (EWPOs) and, in case of the IDM, from DM observables, will be discussed separately below. We further want to mention that limits on 
the total decay width of the $125\gev$ Higgs boson (with current value $\Gamma_\text{tot}\leq13~\MeV$~\cite{Khachatryan:2016ctc}), as well as the total decay widths of the electroweak gauge bosons, could give rise to additional constraints.
Moreover, in the IDM, {restrictions} arise from long-lived charged particle searches at the LHC in the case of almost degenerate masses between the inert charged scalar boson and the DM candidate, {discussed in detail in Ref.~\refcite{Belyaev:2016lok}}. Further model-{specific} limits are discussed in Ref.~\refcite{Ilnicka:2015jba}.

\subsection{Electroweak precision observables}
\label{Sect:EWPOs}
A standard way to {check constraints from electroweak precision observables (EWPOs) in new physics models is to entertain the so-called} oblique parameters, where the most relevant variables are the $S,\,T,\,U$ parameters.\cite{\oblique} The constraints from $S$, $T$ and $U$ can be implemented by evaluating
\begin{align}
\chi^2_\mathrm{STU} = \mathbf{x}^T \mathbf{C}^{-1} \mathbf{x},
\end{align}
with $\mathbf{x}^T = (S - \hat{S}, T - \hat{T}, U - \hat{U})$, where the central values $\hat{S},\hat{T},\hat{U}$, as well as the covariance matrix $\mathbf{C}$ are provided by a global fit to the EWPOs.\cite{Baak:2014ora,Arbuzov:2005ma} Requiring $\chi^2_\mathrm{STU} \le 8.025$ corresponds to a maximal $2\sigma$ deviation given the three {statistical} degrees of freedom.

\begin{figure}[b!]
\centering
\begin{minipage}{0.49\textwidth}
\includegraphics[width=\textwidth]{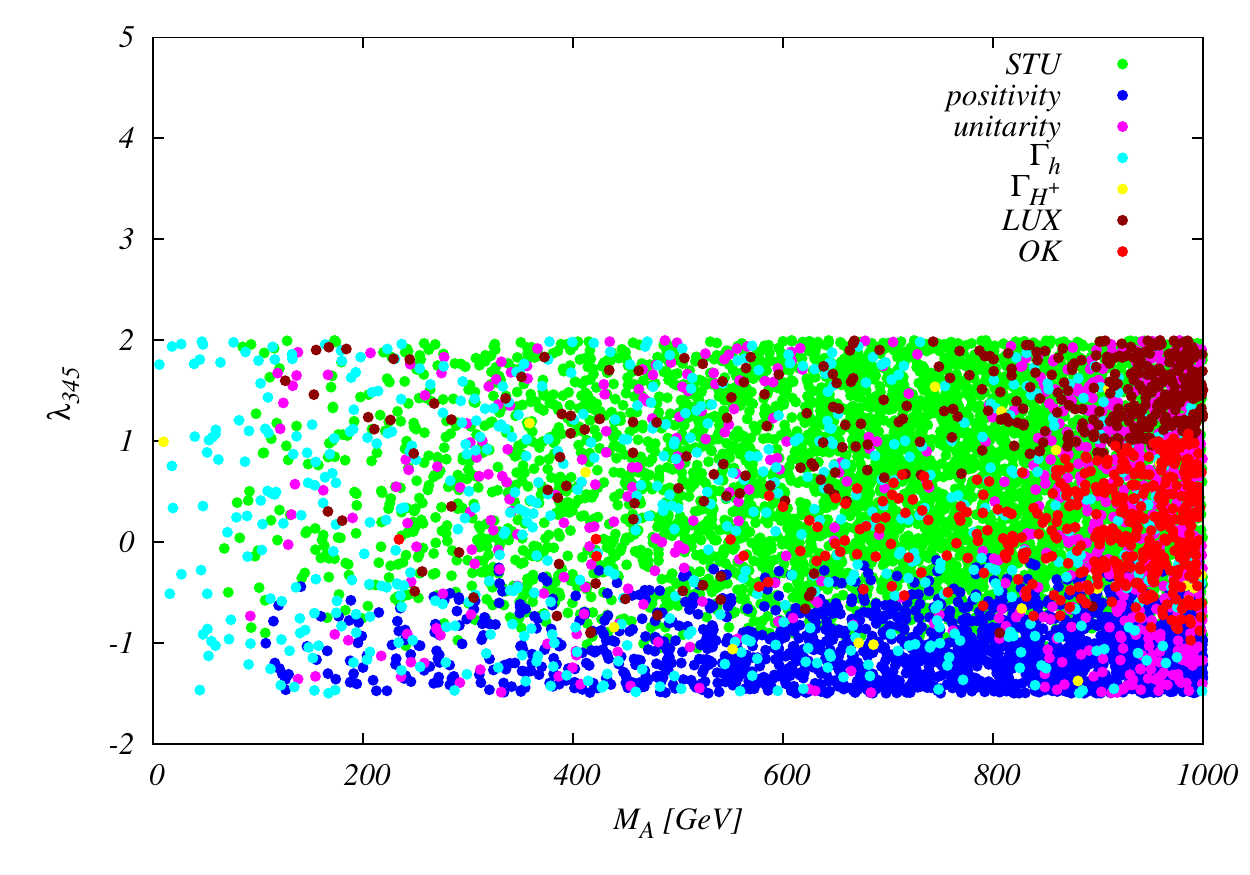}
\end{minipage}
\begin{minipage}{0.49\textwidth}
\includegraphics[width=\textwidth]{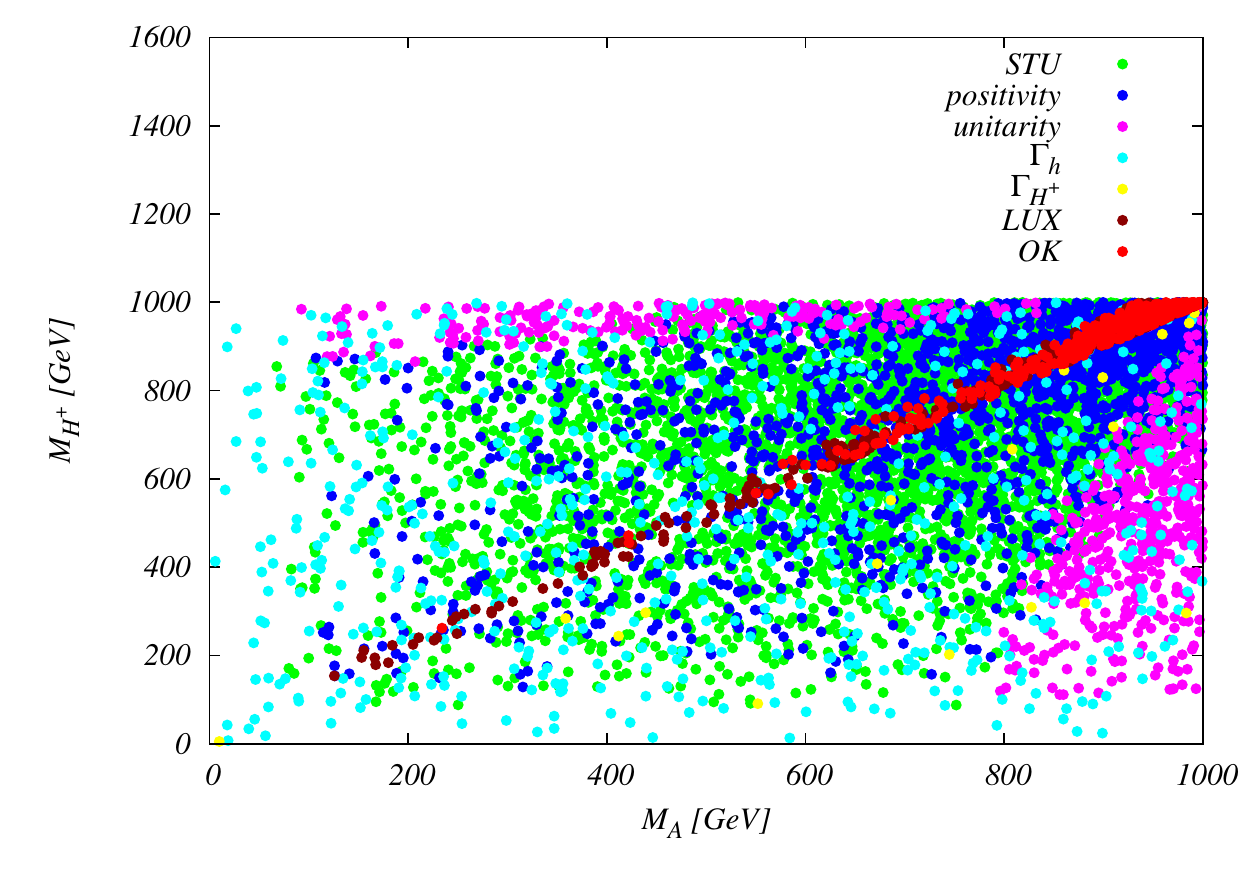}
\end{minipage}
\caption{\label{fig:idm_all} Combined constraints in two exemplary {2D parameter projections} in the Inert Doublet Model. The colored points represent parameter points that were excluded by the individual constraints, {which} are applied in the same order as named in the legend. Allowed points are displayed in red; \emph{left:} ($M_A,~\lam_{345}$) plane, \emph{right:} ($M_A, M_{H^\pm}$) plane. Figure taken from Ref.~\protect\refcite{Ilnicka:2015jba}.}
\end{figure}

We have applied this procedure to both models. In the real singlet extension, however, the derived bound from EWPOs are always superseded by other stronger constraints. In the IDM, as can be inferred from Fig.~\ref{fig:idm_all}, the EWPOs do impose notable constraints in the parameter space. However, {a clear separation of allowed and excluded regions in the shown two-dimensional projected parameter planes cannot be identified.}

\begin{figure}[t!]
\begin{center}
\begin{minipage}{0.49\textwidth}
\begin{center}
\includegraphics[width=\textwidth]{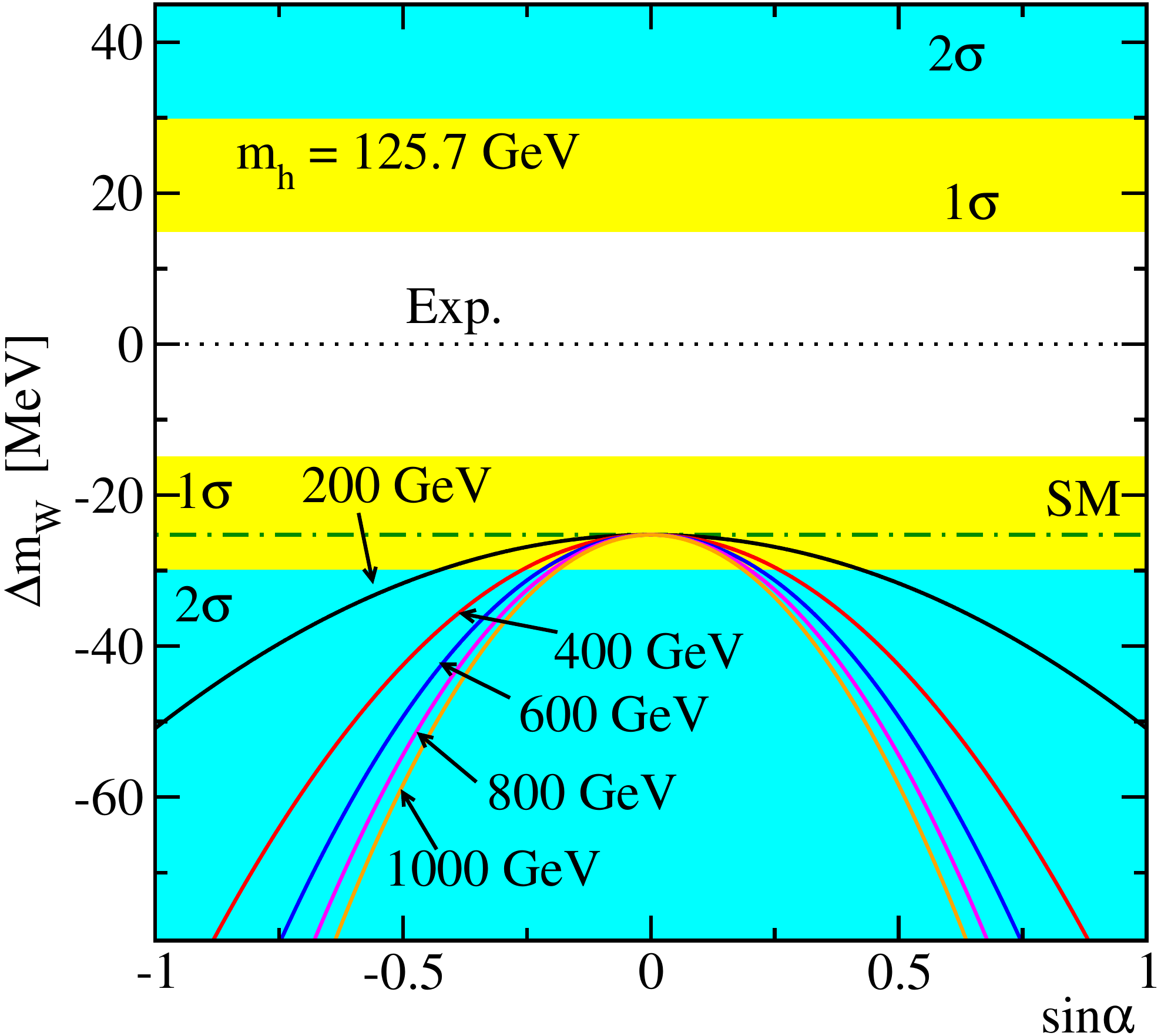}\\
{\tiny $m_h\,=\,125.7\,\GeV$}
\end{center}
\end{minipage}
\begin{minipage}{0.49\textwidth}
\begin{center}
\includegraphics[width=\textwidth]{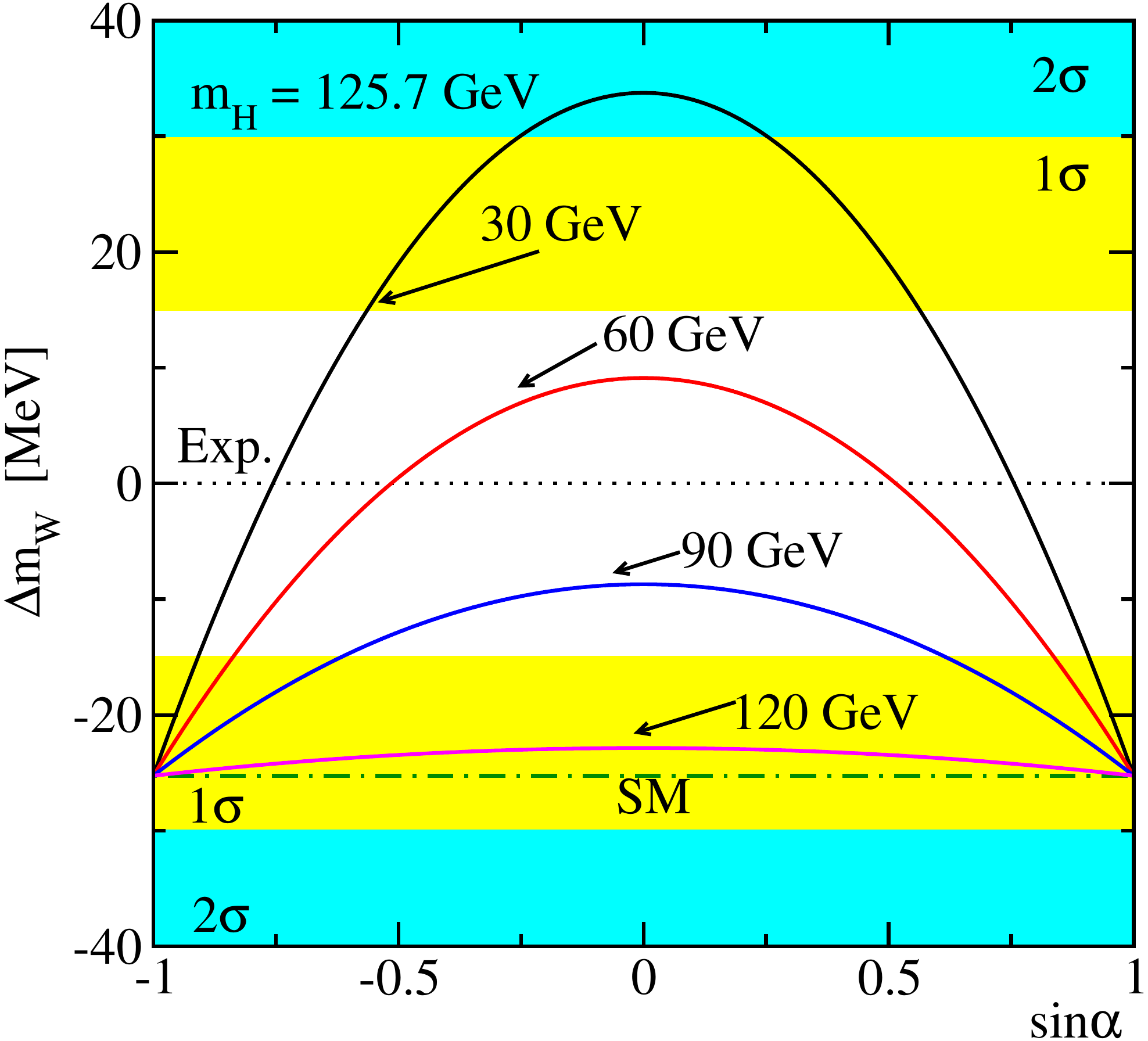}\\
{\tiny $m_H\,=\,125.7\,\GeV$ }
\end{center}
\end{minipage}
\caption{\label{fig:mw_dep} Dependence of $\Delta m_W$, as defined in Eq.~\eqref{eqn:dmw}, on the mixing angle $\sin\alpha$ for cases where the lighter {\sl(left)} or heavier {\sl (right)} scalar boson corresponds to the $125\gev$ Higgs state, for different values of the additional scalar boson mass. Figure taken from Ref.~\protect\refcite{Lopez-Val:2014jva}.}
\end{center}
\end{figure}

{Another way} to constrain new physics models from the electroweak sector is to calculate specific observables that are highly sensitive to new physics contributions. A prime example for this is the mass of the $W$-boson. Here, the current experimental world-average value is given by\cite{Patrignani:2016xqp}
\begin{\eqn}
m^\text{exp}_W\,=\,(80.315\,\pm\,0.015)\,\GeV.
\label{Eq:mWexp}
\end{\eqn}
In Ref.~\refcite{Lopez-Val:2014jva}, the consequences of using the $W$-boson mass as a stand-alone constraint were investigated {in the {framework of the} real singlet extension}. For the case where $m_H\,\geq\,125\,\GeV$, this indeed leads to strong limits on the allowed range for $\sin\al$ (see Fig.~\ref{fig:singlet_constraints}). The main reason for the strength of this constraint is the current discrepancy between the experimental value, Eq.~\eqref{Eq:mWexp}, and the SM prediction of $80.360\,\GeV$.\cite{Awramik:2003rn} The {$\sin\alpha$} dependence of {the discrepancy,}
\begin{\eqn}\label{eqn:dmw}
\Delta\,m_W\,\equiv\,m_W^\text{singlet}-m_W^\text{exp},
\end{\eqn}
is shown in Fig.~\ref{fig:mw_dep} for various values of the additional scalar boson mass. In the scenario where $m_h \simeq 125\gev$, higher-order electroweak corrections increase the discrepancy between the {BSM} prediction and the experimental value, therefore leading to strong constraints when requiring agreement with the experimental value on a $2\sigma$ level. On the other hand, for $m_H\simeq125\gev$, the corrections are able to reconcile the measurement with the theory prediction in certain regions of the parameter space. However, in this case Higgs signal rate measurements and {direct searches at LEP} strongly constrain the mixing angle to values $|\sin\al|\approx 1$.{\cite{Robens:2015gla,Robens:2016xkb}}

{In summary,} electroweak precision observables pose important constraints on new physics scenarios with extended scalar sectors. {In particular,} in cases where they further increase the discrepancy between SM prediction and observed value for specific observables, as e.g.~the $W$-boson mass, they can render the most stringent limits in large regions of parameter space.

\subsection{Dark matter constraints}
\label{sect:DM}
For models with a dark matter candidate, constraints from DM observables need to be taken into account (see e.g.~Ref.~\refcite{Profumo:2017hqp}). The most important are the following: 
\begin{itemize}
\item The DM relic density produced by the model should not overclose the Universe. This leads to requiring {an upper limit for the cold DM relic abundance of}
\begin{\eqn}\label{eq:planck_up}
\Omega_c\,h^2\,\leq\, 0.1241,
\end{\eqn}
corresponding to the $2\sigma$ allowed upper value from the Planck collaboration\cite{Ade:2015xua}. Note that, by imposing only the upper limit on $\Omega_c h^2$, we do allow for DM to be under-abundant.
\item Limits from {DM} direct detection experiments. In Ref.~\refcite{Ilnicka:2015jba} values from the LUX experiment\cite{Akerib:2013tjd} were {taken into account}; {results using updated experimental limits from LUX}\cite{Akerib:2016vxi} were presented in Ref.~\refcite{Ilnicka:2017gab}. In this review, we present {updated} results using {the latest} constraints from the Xenon1T experiment\cite{Aprile:2017iyp}.\footnote{We employ the digitized data from Ref.~\refcite{PhenoData} in this work.}
\item Limits from DM indirect detection experiments. Most recent bounds stem from SuperKamiokande (see e.g. Refs.~\refcite{dmwww,Choi:2015ara}), Icecube\cite{Aartsen:2017nbu}, and Fermi-LAT\cite{Fermi-LAT:2016uux} for $b\bar{b},\,\tau\,\tau$,  $W^+\,W^-$, and photonic final states, respectively. We did not include these in the results presented here.\footnote{See e.g.~Ref.~\refcite{Eiteneuer:2017hoh} for a recent fit study incorporating {in}direct detection constraints.}
\end{itemize}
{Predictions} for the DM observables {in the IDM }were obtained using micrOMEGAs\cite{Belanger:2013oya}, {assuming a standard cosmological thermal history.}

{The impact of the} DM constraints {is quite different for} DM candidates with high and low mass values.\footnote{Many of the results discussed here are IDM specific. However, they serve to demonstrate the complementarity of collider searches and astrophysical {observations}, and can be instructive for the investigation of other BSM dark matter scenarios, especially with a Higgs portal to the DM sector.} We will therefore discuss them separately.

\begin{figure}[h]
\centering
\includegraphics[width=0.67\textwidth]{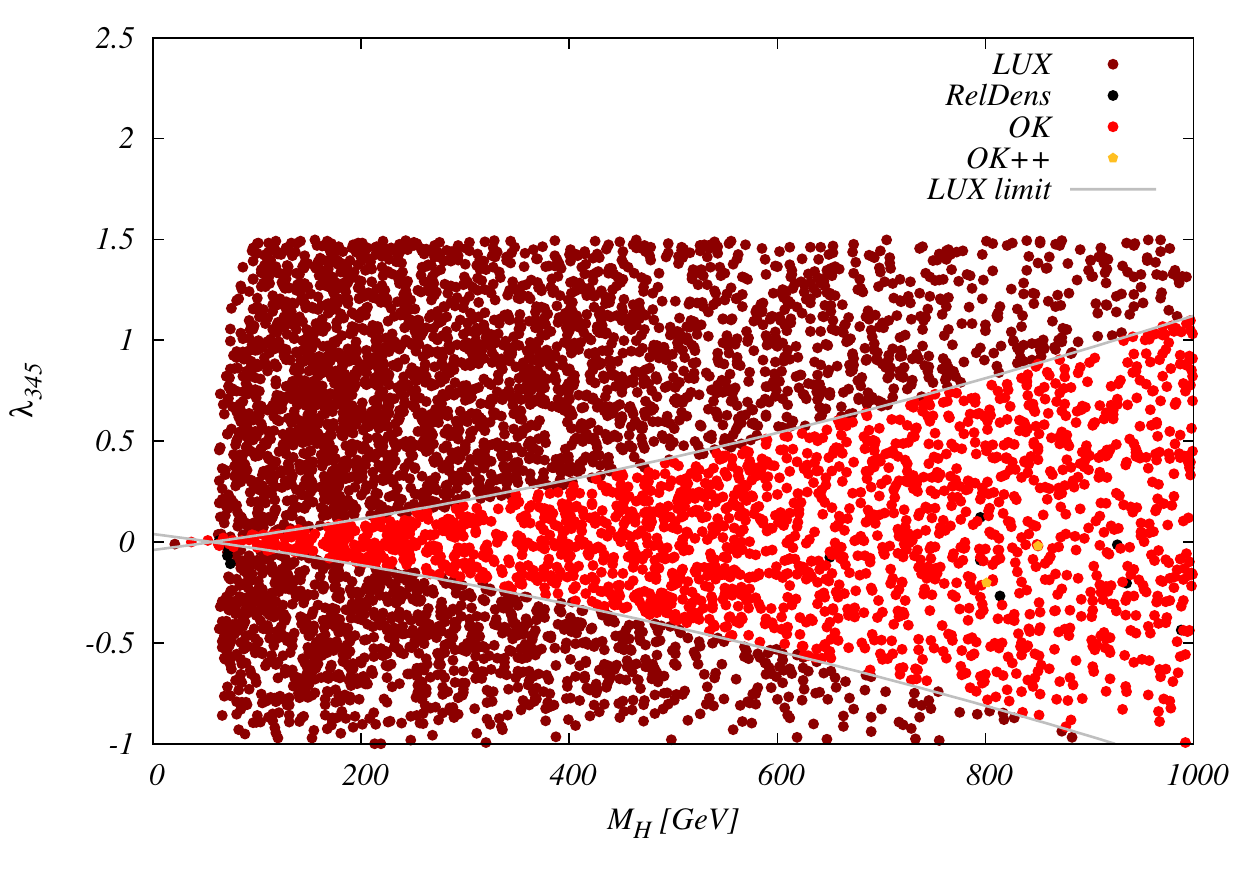}
\caption{ \label{fig:oldluxlim} 
{Constraints on the Inert Doublet Model} in the $(M_H, \lam_{345})$ plane from {DM direct detection experiments}: {Bright red points are allowed and dark red points are excluded.} Additionally, points that reproduce the observed relic density are marked in gold (OK++). Figure taken from Ref.~\protect\refcite{Ilnicka:2015jba}. 
}
\end{figure}

\begin{center}
\begin{figure}
\begin{minipage}{0.49\textwidth}
\includegraphics[width=\textwidth]{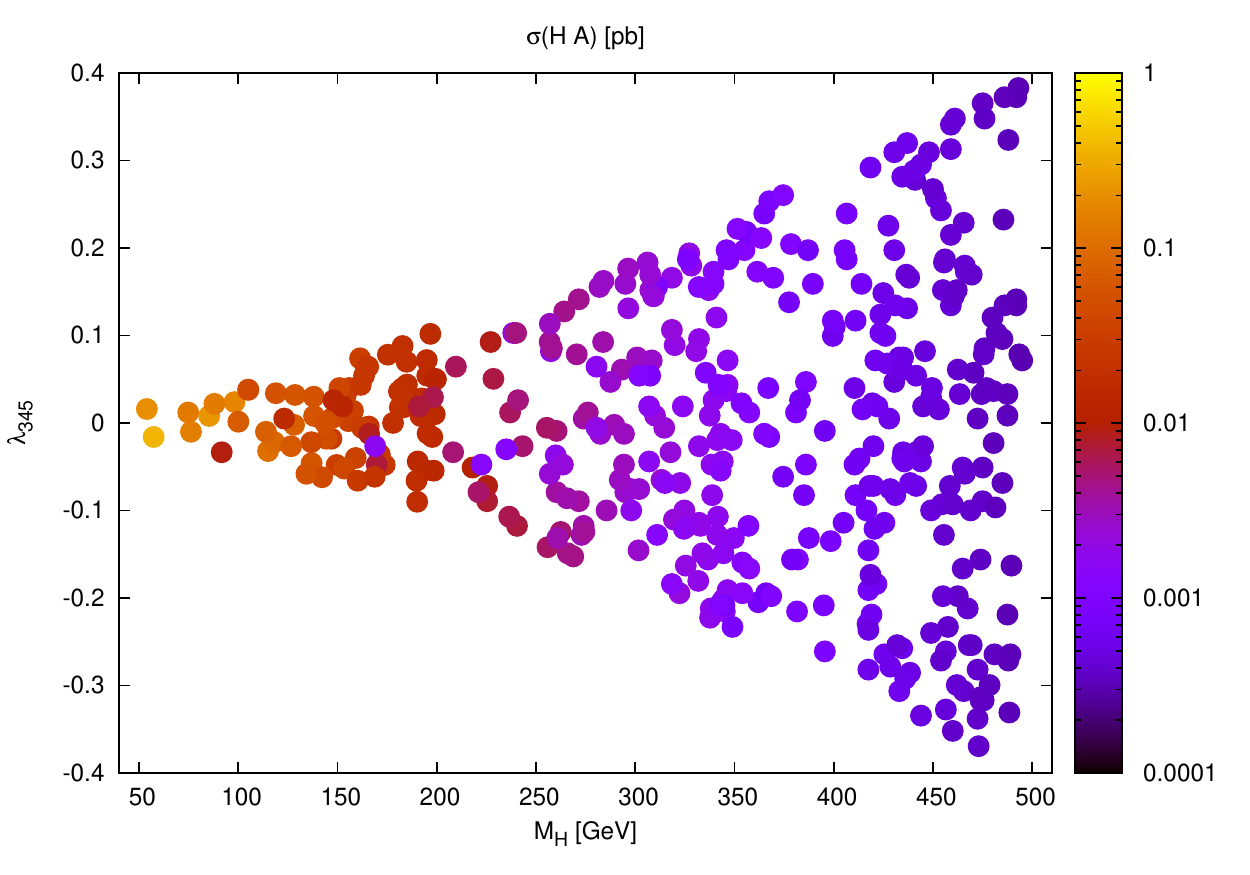}
\end{minipage}
\begin{minipage}{0.49\textwidth}
\includegraphics[width=\textwidth]{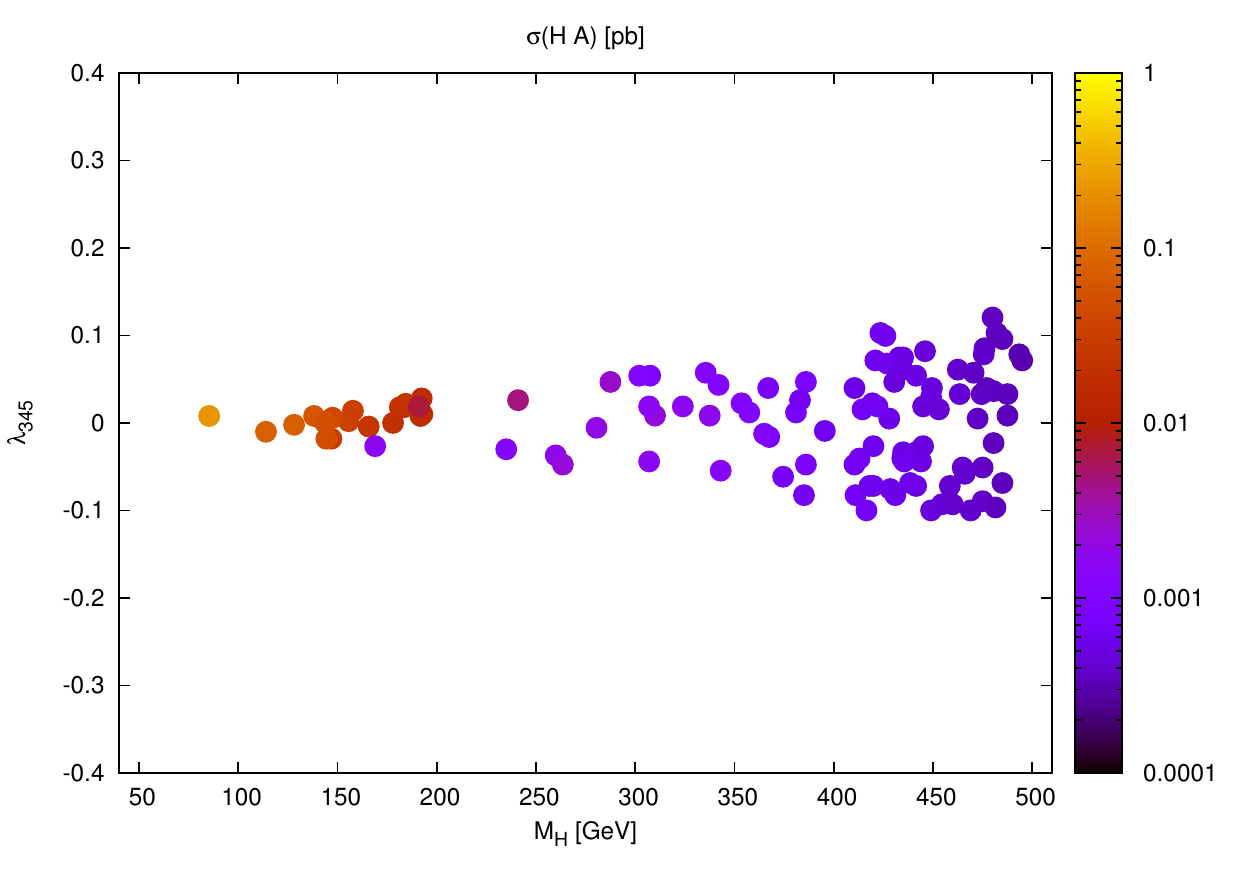}
\end{minipage}
\caption{\label{fig:xsec_change} Production cross section (in \pb) for {$pp\to H A$} production at the 13 \TeV~LHC in the Inert Doublet Model, after applying the LUX {\sl (left)} and XENON1T {\sl (right)} DM direct detection constraints. The allowed parameter space shrinks considerably once the newer experimental bounds are taken into account. The left figure is taken from Ref.~\protect\refcite{Ilnicka:2015jba}.}
\end{figure}
\end{center}

In general, the two {IDM} parameters which are most sensitive to limits from direct detection experiments are the dark matter mass, $M_H$, as well as the coupling $\lam_{345}$.
{As an example, Fig.~\ref{fig:oldluxlim} show the constraints on these parameters (Fig.~taken from Ref.~\refcite{Ilnicka:2015jba}).} It illustrates that in particular for the mass range $M_H\,\gtrsim\,M_h/2$ the LUX limit clearly separates the parameter space. Newer direct detection results improve the limits by an order of magnitude. This is exemplarily shown in Fig.~\ref{fig:xsec_change}, displaying the production cross section for $pp \to HA$ --- one of the dominant production modes of the IDM at hadron colliders\cite{Ilnicka:2015jba,deFlorian:2016spz} --- at the $13~\TeV$ LHC for regions allowed after applying LUX {\sl (left)} and {2017} XENON1T {\sl(right)} DM direct detection limits. {These updated} constraints lead to a {significant} decrease of the allowed parameter regions. The production cross sections were calculated using \texttt{Madgraph-5}\cite{Alwall:2011uj}, using a UFO model file for the IDM from Ref.~\refcite{Goudelis:2013uca}.

For low DM masses, in particular the case $M_H\,\leq\,M_h/2$, constraints from the LHC Higgs signal strengths play an important role{, as the new decay mode $h\to HH$ of the observed Higgs boson becomes kinematically accessible {(see Sect.~\ref{sect:higgssignal})}}. However, as can be seen from Fig.~\ref{fig:lowmhidm_luxii}, the most stringent bounds in this region stem from relic density constraint, Eq.~\eqref{eq:planck_up}. {Here,} large relic density values are obtained from the predominant annihilation channels $HH\rightarrow b\bar{b}$ and $HH\rightarrow hh$, see Refs.~\refcite{Goudelis:2013uca,Ilnicka:2015jba,danandme}. The resulting parameter space, {after including the latest} XENON1T limits, is shown in Fig.~\ref{fig:low_xenon}. {The DM relic density limit prevails in most regions of the parameter space; {in combination with direct detection limits, this leads} to $|\lam_{345}|\,\lesssim\,0.01$ for $M_H\,\leq\,M_h/2$. Let us emphasise, that the gain in sensitivity between LUX and XENON1T lead to a decrease of allowed values for this coupling by a factor two with respect to the results presented in Ref.~\refcite{Ilnicka:2015jba}. 

As discussed above, astrophysical constraints mainly affect the coupling $\lam_{345}$ in the IDM.  On the other hand, production and decay processes at the LHC are dominantly determined by the masses of the inert scalars and the SM electroweak couplings.\cite{Ilnicka:2015jba} Therefore, the IDM {is} a prime example for a model that {features an important complementarity between dark matter and collider searches}. In fact, recent developments {in the field of} DM direct detection experiments will hopefully allow to further constrain {or discover} this model in the near future.\cite{Battaglieri:2017aum}

\begin{figure}
\begin{center}
\includegraphics[width=0.8\textwidth]{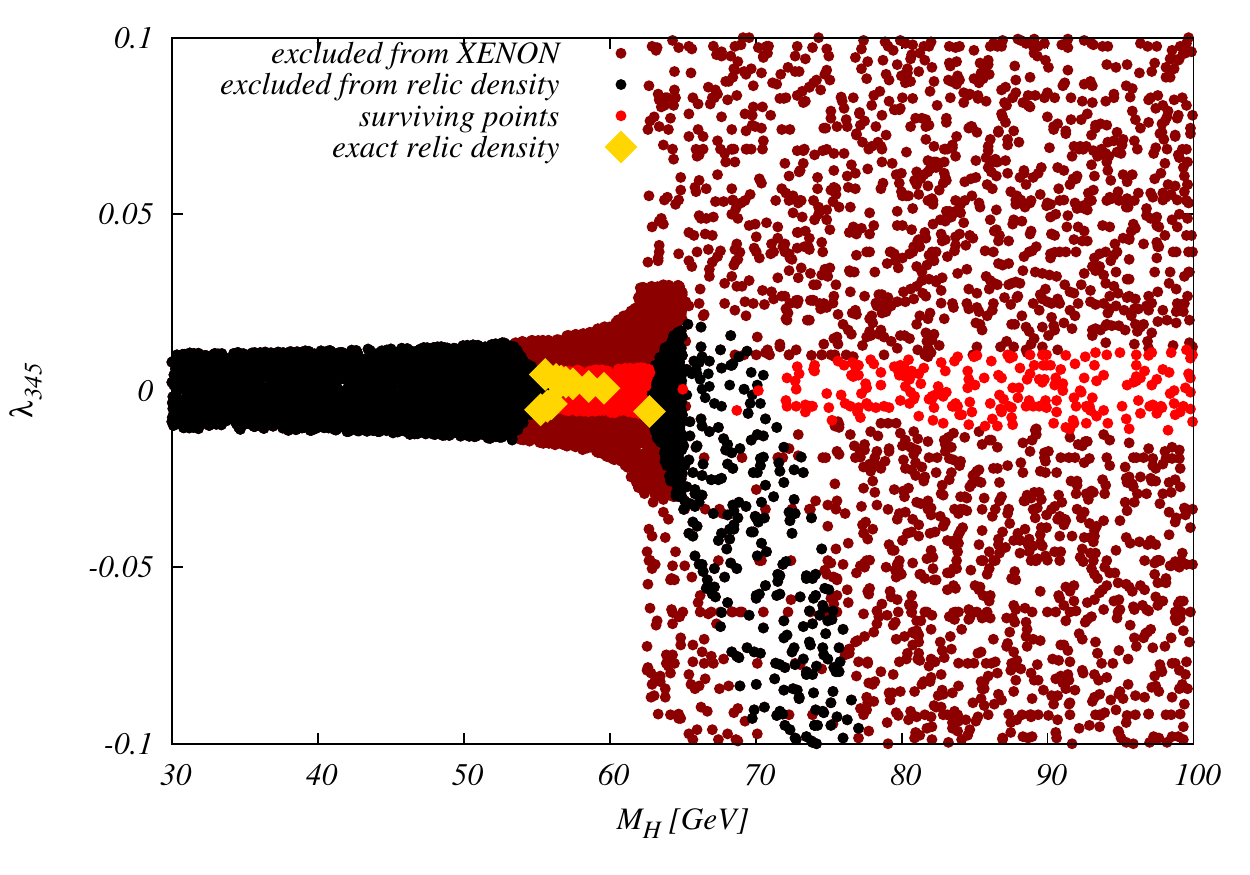}
\caption{\label{fig:low_xenon} {Comparison of dark matter constraints in the Inert Doublet Model in the ($M_H, \lam_{345})$ parameter plane for low DM masses. Parameter points excluded by the latest XENON1T limit or the DM relic density are shown in \emph{brown} and \emph{black}, respectively. Allowed points are shown in \emph{red}, and \emph{yellow} points correspond to scenarios that exactly reproduce the DM relic density within $95\,\%$ C.L..}}
\end{center}
\end{figure}

\section{Conclusions}
We briefly reviewed theoretical and experimental constraints that need to be taken into account when considering models with extended scalar sectors. For {illustration}, we discussed two simple {Beyond the Standard Model} scenarios: {First, the real scalar singlet extension of the Standard Model, and second, the Inert Doublet Model, i.e.}~a two Higgs doublet model featuring a dark matter candidate. {While direct LHC searches for additional scalar states, as well as} the measurement of the $125\gev$ Higgs boson {signal strengths, limit} the models' parameter space significantly, we also pointed to the importance of complimentary limits {from other areas}, with a focus on electroweak precision observables and dark matter {constraints}. In summary, {BSM scenarios with extended scalar sectors can be probed in various ways,} {and we hope that the near future will bring a discovery either directly in collider or dark matter searches, or indirectly through precision measurements (or both).}

\section*{Acknowledgments}
TR wants to thank G.M. Pruna, D. Lopez-Val, and G. Chalons for fruitful collaboration on related topics. This material is based in part upon work supported by part by the National Science Centre, Poland, the HARMONIA project under contract UMO-2015/18/M/ST2/00518 (2016-2019), the National Science Foundation under Grant No. 1519045, by Michigan State University through computational resources provided by the Institute for Cyber-Enabled Research, and by grant K 125105 of the National Research, Development and Innovation Fund in Hungary.
{AI is supported by the 7th Framework Programme of the European Commission through the Initial Training Network HiggsTools PITN-GA-2012-316704. TS acknowledges support from the DESY Fellowship programme.}

\end{document}